\documentclass[epj,twocolumn]{svjour}

\usepackage{color}
\usepackage{amssymb}
\usepackage{amsmath}
\usepackage{epsfig}
\usepackage{graphics}
\usepackage{graphicx}
\usepackage{dcolumn}
\usepackage{subfigure}
\usepackage{bm}

\def\be{\begin{equation}}
\def\ee{\end{equation}}

\begin{document}

\title{Dynamics in quantum Ising chain driven by 
inhomogeneous transverse magnetization}

\author{Sirshendu Bhattacharyya\inst{1,}\inst{2,}\thanks{\email{sirs.bh@gmail.com}}
\and Subinay Dasgupta\inst{2}}
\institute{Department of Physics, R.R.R Mahavidyalaya,
 Radhanagar, Hooghly 712406, India 
 \and Department of Physics, University of Calcutta,
  92 Acharya Prafulla Chandra Road, Kolkata 700009, India}


\abstract{
We study the dynamics caused by transport of transverse magnetization in one dimensional transverse Ising chain at zero temperature. We observe that a class of initial states having product structure in fermionic momentum-space and satisfying certain criteria, produce spatial variation in transverse magnetization. Starting from such a state, we obtain the transverse magnetization analytically and then observe its dynamics in presence of a homogeneous constant field $\Gamma$. In contradiction with general expectation, whatever be the strength of the field, the magnetization of the system does not become homogeneous even after infinite time. At each site, the dynamics is associated with oscillations having two different timescales. The envelope of the larger timescale oscillation decays algebraically with an exponent which is invariant for all such special initial states. The frequency of this oscillation varies differently with external field in ordered and disordered phases. The local magnetization after infinite time also characterizes the quantum phase transition.
}

\titlerunning{Dynamics in quantum Ising chain}
\authorrunning{S. Bhattacharyya, S. Dasgupta}
\maketitle
\section{Introduction}
 
Quantum many-body system out of equilibrium has been subjected to upsurging interest in recent times. Extensive studies have been done over decades giving rise to many new findings in this field \cite{Pol}. Lots of questions still remain unresolved too. From the results obtained so far, it is impossible to draw a generic picture of the dynamics out of equilibrium. In this context, integrability has emerged to be a very crucial perspective \cite{Rigol-Nature,Rigol-PRL,Kormos,Collura}. Integrable and non-integrable systems generally have different response in such dynamics. However, a complete picture is yet to be drawn on the role of integrability in nonequilibrium phenomena. \\

Integrable systems with many local conserved quantities have restricted dynamics and unlike non-integrable systems, their relaxation process is dependent on the initial state. Non-integrable systems, on the other hand, were reported to exhibit thermalization in some earlier works \cite{Biroli}. Evidences however came into scenario where some non-integrable systems are shown to have weak thermalization or even no thermalization \cite{Bauls,Kinoshita}. Numerical studies on non-integrable systems has been done recently in this regard \cite{Blab}. Moreover, it has been shown \cite{Suzuki} that in a completely integrable system, apart from the usual non-thermal behavior of some quantities, there do exist some other quantities that exhibit thermal behavior which is counterintuitive to the prevalent ideas on thermalization \cite{Rigol-Nature,Rigol-PRL}. Owing to some more elaborate studies, \cite{Eisert,Essler} the role of integrability on the behavior of response functions have been rendered more debatable. Another important question is whether their lies any generic relation between integrability and transport properties. Efforts have been given in that part too by some preceding works \cite{Castella,Zotos}.\\

Emergence of current of a physical quantity generally renders a system to a nonequilibrium one. Transport of that quantity due to presence of current thereby gets attention as an important aspect of nonequilibrium dynamics. Transport properties, especially magnetization transport in low dimensional spin systems have received attention of the experimentalists too gaining importance in the area of spintronics, nano-device applications \cite{Wolf,Meier,Ramanathan}. Earlier theoretical studies \cite{Rieder,Saito} have confirmed the fact that the properties of transportation in closed quantum system differ significantly from its classical counterpart. A number of works, both analytic and numerical, have been done on introducing current of a quantity in the integrable Hamiltonian to observe its relaxation dynamics \cite{Antal1,Antal2,Antal3,Santos,Divakaran}. In these papers, current-carrying states have been generated in the system in two ways, either by adding current to the Hamiltonian or by preparing an inhomogeneous initial state. Introduction of magnetization current 
results in oscillatory nature and power-law decay of correlations in space. A typical initial state with a steplike inhomogeneity in magnetization even yields a scaling form of the same in the large-time limit. Analytic as well as simulative studies have also been made on nonequilibrium transport of magnetization in open quantum spin chain \cite{Znidaric,Banerjee}. These studies have been able to shed some light upon the behavior of physical quantities in an integrable system while trying to equilibrate.\\

A scattered context like this has been the motivation of more investigation in relaxation dynamics of integrable quantum system generally followed by a quench \cite{Essler,Piroli,Caux}. In this article, we study the dynamics due to transport of transverse magnetization in a quantum Ising chain. Transportation of magnetization is induced by preparing an initial state which produces sinusoidally inhomogeneous transverse magnetization and thus puts the system into a nonequilibrium state. We let it evolve in presence of a homogeneous and constant (in time) transverse field with a view to studying whether the transport properties finally make the magnetization homogeneous or not and how the magnetization relaxes with time. Somewhat similar study has been made on XXZ spin chain showing oscillatory behavior and power-law decay \cite{Lancaster}. Power-law relaxation of local magnetization has been observed in Heisenberg spin chain too \cite{Deguchi}. Study of temporal behavior of transverse magnetization from a thermally inhomogeneous state has also been done in quantum XX chain \cite{Platini}. In our case, to prepare such an initial configuration,  
a new protocol is introduced : We include two odd-occupation basis of zero eigenvalue in the initial state which have product structure in fermionic momentum space. Our observation in this case is that the later dynamics driven by the external transverse field cannot make the magnetization homogeneous even after infinite time. This phenomena is surprising as our classical intuition demands the system to wipe out its inhomogeneity in course of time under strong external field. That is how the quantum behavior of a system, especially of an integrable one, counters our prevailing (classical) idea of a physical process. Moreover, the underlying dynamics at each site is shown to possess two different timescales of oscillation. The smaller timescale contributes insignificant undulations whereas the oscillations of larger timescale exhibit a generic power-law decay. The exponent of the decay is proved to be independent of the strength of the external field. This type of decay also indicates absence of thermalization, in agreement with previous works \cite{Antal3,Hunyadi}, whereas in contrast with that work, no scaling form is observed in the large-time limit for this type of initial configuration. A very recent work on thermalization also supports the fact that such type of decay indicates lack of thermalization \cite{Tavora}.\\

 Another important feature of our work is that the phenomena of quantum phase transition is manifested in the dynamics in two ways as can be established analytically, (i) the frequency of the characteristic oscillation is a monotonic function of the external field as long as the system is in the ordered phase and becomes independent of the field when it crosses the critical value to the disordered phase, (ii) the transverse magnetization after infinite time at each site shows different behavior as a function of external field in the ordered and disordered phase and thus produces nonanalyticity at the critical point.\\
 
  Lastly, our initial configuration contains several parameters and the key features of the dynamics we observe, are true for arbitrary values of those parameters. Thus, in spite of the fact that the Hamiltonian is integrable, some characteristics of the dynamics is valid for all the initial configurations satisfying a special criteria.\\ 

In the next section we give detailed description of the model and of the initial state. Section III contains the exact analytic treatment of transverse magnetization and its dynamics. In the last section, conclusions are drawn from the results obtained along with general discussions.

\section{The Model and the Initial Configuration}

One dimensional spin-$\frac{1}{2}$ transverse-field Ising chain of $N$ sites is described by the Hamiltonian \cite{Pfeuty}
\be   \mathcal{H}= -\sum_{i=1}^{N} s^{x}_{i} s^{x}_{i+1} - \Gamma\sum_{i=1}^{N} s^{z}_{i}    \ee
where $\Gamma$, scaled by the coupling constant, is the external transverse field and $s^{x,z}$ are two components of Pauli spin matrices. This Hamiltonian shows an ordered (ferromagnetic) to disordered (paramagnetic) phase transition at the critical point $\Gamma = 1$.\\

The well-known Jordan-Wigner transformation enables us to transform the Hamiltonian into a direct sum of commuting Hamiltonians ($\mathcal{H}_k$) of nonlocal free fermions of momenta $\pm k$ \cite{LSM,Damski,BKC-book}
\begin{eqnarray}
 \mathcal{H}_k &=& (-2 i \sin k) \left[ a_k^{\dagger}a_{-k}^{\dagger} + a_k a_{-k} \right] \nonumber \\
&&- 2(\Gamma + \cos k) \left[ a_k^{\dagger}a_k + a_{-k}^{\dagger}a_{-k} - 1 \right] \label{Hk_def} 
\end{eqnarray} 
where $a^{\dagger}_k$ and $a_k$ are fermionic creation and annihilation operator in momentum space and $k=(2\ell+1)\pi/N$, with $\ell=0,1,\cdots,\frac{N}{2}-1$.
The operator $\mathcal{H}_k$ has four eigenstates. The even-occupation eigenstates of $\mathcal{H}_k$ are spanned by two basis states namely, $|0,0 \rangle_k$ and $|1,1 \rangle_k$ where the numbers signify the occupation number of the fermions having momenta $+k$ and $-k$ respectively. We denote the eigenstates of $\mathcal{H}_k$ within this subspace as $|(\Gamma ,k)_{\mp}\rangle$ with eigenvalues $\mp\lambda_k$, where
\begin{eqnarray}
\lambda_k & = & 2\sqrt{\Gamma^2 +1 +2\Gamma \cos k} \\
|(\Gamma ,k)_{-}\rangle & = & i \cos \theta |1,1\rangle_k - \sin \theta |0,0\rangle_k \\
|(\Gamma ,k)_{+}\rangle & = & i \sin \theta |1,1\rangle_k + \cos \theta |0,0\rangle_k 
\end{eqnarray}
with $\tan \theta_k = -\sin k /\left[ \Gamma + \cos k + \sqrt{\Gamma^2 +1 +2\Gamma \cos k}\; \right]$.

The other basis states $|0,1\rangle_k$ and $|1,0\rangle_k$ of $\mathcal{H}_k$ have zero eigenvalue.
For any wave function $|\Psi\rangle$, the transverse magnetization at the $n$-th site is given by 

\begin{equation}
M_z = \dfrac{2}{N}\sum_{k_1,k_2=-\pi}^{\pi} e^{i(k_2-k_1)n} \langle \Psi | a^{\dagger}_{k_1} a_{k_2} |\Psi\rangle \;-\;1
\label{Mnz}   \end{equation}
It is important to observe that this magnetization is {\it homogeneous} if $|\Psi\rangle$ is the ground state or any other state comprising of $|00\rangle_k$ and $|11\rangle_k$.\\

We construct an initial configuration incorporating the odd-occupation basis ($|01\rangle_k$ and $|10\rangle_k$) in momentum space  alongwith the states $|00\rangle_k$ and $|11\rangle_k$. We choose
\be    |\Psi(0)\rangle = \bigotimes |\Psi_k\rangle \label{def_psi}  \ee 
with  $|\Psi_k\rangle = \alpha_k |11\rangle_k + \beta_k |00\rangle_k + \gamma_k |10\rangle_k + \gamma_k |01\rangle_k$.
The normalization condition requires 
\be |\alpha_k|^2 + |\beta_k|^2 + 2|\gamma_k|^2 = 1 \label{normalization} \ee
 for all $k$. One may note that the newly constructed state is not an eigenstate of the fermionic Hamiltonian $\mathcal{H}_k$. We shall now show that one can calculate analytically the magnetization at any given site for this initial state under certain conditions. The expression for magnetization Eq. (\ref{Mnz}) involves jump of fermions from any site $k_2$ to any site $k_1$. Let us first assume that $k_1$ and $k_2$ are both positive and $k_1 < k_2$. For such a jump, the sign of the resulting term is then determined by whether the fermion crosses an even or odd number of fermions during its flight. 

\begin{eqnarray}& &  a_{k_1}^{\dagger} a_{k_2} \left(\Pi_{k=k_1}^{k_2}  |\Psi_k\rangle \right)
= \pm  \left(\gamma_{k_1} |11\rangle_{k_1} + \beta_{k_1}  |10\rangle_{k_1} \right)   \nonumber \\ 
& &  \left[ \Pi_{k>k_1, k< k_2}  \left(\alpha_k |11\rangle_k + \beta_k |00\rangle_k + \gamma_k |10\rangle_k + \gamma_k |01\rangle_k
 \right)\right] \nonumber \\ 
 & &  \left(\alpha_{k_2} |01\rangle_{k_2} + \gamma_{k_2} |00\rangle_{k_2} \right)     \end{eqnarray}

The sign will be plus (minus) if the number of fermions in the range $k_1 < k < k_2$ for a given term in the expanded form of the portion within square brackets is even (odd). However, one observes that, the correct sign is obtained from the equality,
\begin{eqnarray}& &  a_{k_1}^{\dagger} a_{k_2} \left(\Pi_{k=k_1}^{k_2}  |\Psi_k\rangle \right)
=   \left(\gamma_{k_1} |11\rangle_{k_1} + \beta_{k_1}  |10\rangle_{k_1} \right)   \nonumber \\ 
& &  \left[ \Pi_{k>k_1, k< k_2}  \left(\alpha_k |11\rangle_k + \beta_k |00\rangle_k - \gamma_k |10\rangle_k - \gamma_k |01\rangle_k
 \right)\right] \nonumber \\ 
 & &  \left(\alpha_{k_2} |01\rangle_{k_2} + \gamma_{k_2} |00\rangle_{k_2} \right)     \label{ak}  \end{eqnarray}

In order to calculate the magnetization $M_z$ at site $n$, we need to operate $ \langle \Psi(0) | $ on Eq. (\ref{ak}). 
However, a simplification occurs if we note that
\begin{eqnarray}& & \left(\alpha_k^* \langle 11| + \beta_k^* \langle 00|_k + \gamma_k^* \langle 10|_k + \gamma_k^* \langle 01|_k \right) \nonumber \\
&& \left(\alpha_k |11\rangle_k + \beta_k |00\rangle_k - \gamma_k |10\rangle_k - \gamma_k |01\rangle_k
 \right) \nonumber \\
 && = |\alpha_k|^2 + |\beta_k|^2 - 2|\gamma_k|^2  \end{eqnarray}

Hence, if we ensure that
 \be |\alpha_k|^2 + |\beta_k|^2 = 2|\gamma_k|^2 \label{zero} \ee  
for all $k$,  the quantity $ \langle \Psi(0) | a_{k_1}^{\dagger} a_{k_2}  |\Psi(0) \rangle $
will be non-zero {\em only when} either $k_1=k_2$ or $k_1$ and $k_2$ are two successive points so that $|k_2 - k_1| = 2\pi/N = u$ (say). Note that the constraints (\ref{normalization}) and (\ref{zero}) need $|\gamma_k|=\frac{1}{2}$ and $|\alpha_k|^2 + |\beta_k|^2 = \frac{1}{2}$ for all $k$. We assume additionally, that $\gamma_k=\frac{1}{2}$ for all $k$. One can now obtain the expression for magnetization $M_z(n,t)$ by considering all possible choices of $k_1$ and $k_2$,
\[ M_z (n,t=0)  =  - \frac{1}{2} + \frac{2}{N}\sum_{k=0}^{\pi} \;m_k \]
where
\begin{eqnarray} 
& &  m_k =  2\alpha^{*}_k \alpha_k \;+\;  \frac{1}{2} \lbrace (\beta_k\beta^{*}_{k+u}-\alpha^{*}_k\alpha_{k+u})\cos un \nonumber \\
 && +\; (\beta_k\alpha_{k+u}-\alpha^{*}_k\beta^{*}_{k+u})i\sin un \nonumber \\
&& +\; (\beta_k\beta^{*}_{k+u}+\alpha^{*}_k\alpha_{k+u})\cos (2k+u)n \nonumber \\
&& +\; (\beta_k\alpha_{k+u}+\alpha^{*}_k\beta^{*}_{k+u})i\sin (2k+u)n \;+\; c.c.\rbrace
\label{Mn0}
\end{eqnarray}
Here the sum over $k$ runs for $(N/2)$ values between $0$ and $\pi$ and $c.c.$ stands for complex conjugate. For large $N$, only the $\sin un$ and $\cos un$ terms contribute to the magnetization at a site.

 \section{Dynamics of Transverse Magnetization}

In order to study the dynamics we first note that, from Eq. (\ref{def_psi}) the wave function at time $t$ may be obtained as
\begin{eqnarray}
|\Psi_k(t)\rangle &=& e^{-i\mathcal{H}_k t}|\Psi_k (t=0)\rangle \nonumber \\ 
&=& \alpha_k^{\prime} |11\rangle + \beta_k^{\prime} |00\rangle + \gamma |10\rangle + \gamma |01\rangle 
\end{eqnarray}
where 
\begin{eqnarray*}
\alpha_k^{\prime} &=& (\alpha_k\cos^2\theta_k e^{i\lambda_k t} + \alpha_k\sin^2\theta_k e^{-i\lambda_k t} \nonumber \\
&& - i\beta_k\sin\theta_k\cos\theta_k e^{i\lambda_k t} + i\beta_k\sin\theta_k\cos\theta_k e^{-i\lambda_k t}) \nonumber \\
\beta_k^{\prime} &=& (i\alpha_k\sin\theta_k\cos\theta_k e^{i\lambda_k t} - i\alpha_k\sin\theta_k\cos\theta_k e^{-i\lambda_k t} \nonumber \\
&& + \beta_k\sin^2\theta_k e^{i\lambda_k t} + \beta_k\cos^2\theta_k e^{-i\lambda_k t})
\end{eqnarray*} 
with $t$ is scaled by $\hbar$. A very crucial point is that the coefficients $\alpha_k^{\prime}$, $\beta_k^{\prime}$ and $\gamma$ also satisfy the condition Eq. (\ref{zero}).

Hence, the magnetization at time $t$ can also be calculated following the previous procedure :

\be M_z(n,t) = -\frac{1}{2} + \dfrac{2}{N}\sum_{k=0}^{\pi} \; m_k   \label{M1}  \ee
where $m_k$ is given by the expression in Eq. (\ref{Mn0}) with $\alpha_k$, $\beta_k$ replaced by $\alpha'_k$, $\beta'_k$. After this replacement one can express $m_k$ as
\begin{eqnarray}
m_k &=& A_k \;+\; B_k \cos(2\lambda_k t + b_k) \nonumber \\
&& +\; C_k(n)  \cos\lbrace 2(\lambda_{k+u}+\lambda_k)t + c_k(n) \rbrace \nonumber \\
&& +\; D_k(n)\cos\lbrace 2(\lambda_{k+u}-\lambda_k)t + d_k(n) \rbrace  \label{mk} 
\end{eqnarray}
The precise expressions for the amplitudes $A_k, B_k, C_k, D_k$ and the phases $b_k$, $c_k$, $d_k$ in terms of $k$, $n$, $\alpha_k$, $\beta_k$ and $\gamma_k$ can be obtained but is not required for our subsequent calculations.\\

From now on we shall be restricted to the case where $\alpha_k$ and $\beta_k$ are real and equal for all $k$, i.e. $\alpha_k=\alpha$ and $\beta_k=\beta$. Then, in the thermodynamic limit, $M_z(n,t)$ can be written as
\begin{equation} 
M_z(n,t) = -\frac{1}{2} + \dfrac{1}{\pi}\int_0^\pi \left[\mathcal{M}_k+m_k(t)+m_k(t/N)\right] \; dk
\label{M_int} 
\end{equation}
where
\begin{eqnarray*}
\mathcal{M}_k &=& 2\alpha^2  + (\beta^2 - \alpha^2)(1 - \frac{1}{2} \cos un) \sin^2 2\theta_k \nonumber \\
&& + \frac{1}{4}\sin 4\theta_k \sin un + \frac{1}{4}\sin^2 2\theta_k \cos 2kn \\
m_k(t)&=& [(\alpha^2-\beta^2)(1-\cos un) \sin^2 2\theta_k \nonumber \\
&& - \frac{1}{4}\sin 4\theta_k \sin un \;]\cos 2\lambda_k t \nonumber \\
&& + [ 2\alpha\beta\sin 2\theta_k(1-\cos un) ] \sin 2\lambda_k t \\
m_k(t/N)&=& \frac{1}{2}(\cos^4\theta_k + \sin^4\theta_k)[2(\beta^2-\alpha^2)\cos un \nonumber \\
&& + \cos 2kn] \cos 2\pi\lambda'_k\frac{t}{N} \nonumber \\
&& -\; 2\alpha\beta\cos 2\theta_k \sin 2kn \sin 2\pi\lambda'_k\frac{t}{N}
\end{eqnarray*}
We have written $\lambda_{k+u}-\lambda_k = \frac{2\pi}{N}\frac{d\lambda_k}{dk}=u\lambda'_k$ as in the thermodynamic limit the difference between consecutive $k$-points become infinitesimally small. The dynamics is characterized by the behavior of transverse magnetization $M_z(n,t)$ at a time $t$ at site $n$, as given by Eqs. (\ref{M_int}). At any given time (including $t=0$) the magnetization is a continuously varying function of $n$.\\

\begin{figure}
\begin{subfigure}
{\includegraphics[clip,width=8cm, angle=0]{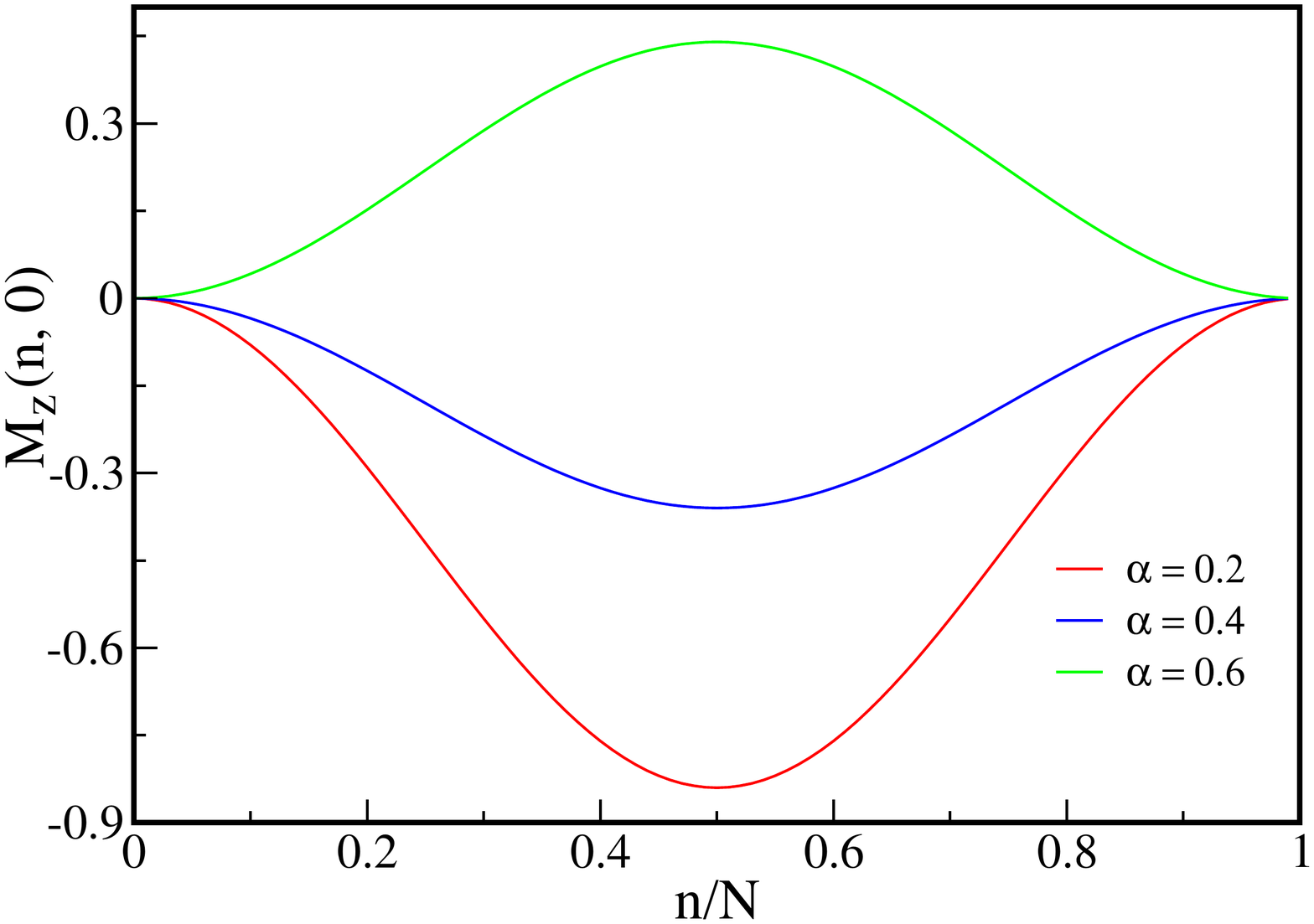}}
\caption{Initial Transverse magnetization as a function of position $n$, for the state (\ref{def_psi}) with different $\alpha$ keeping $\Gamma=0.5$.}
\label{mz0-plot}
\end{subfigure}
\begin{subfigure}
{\includegraphics[clip,width=8cm, angle=0]{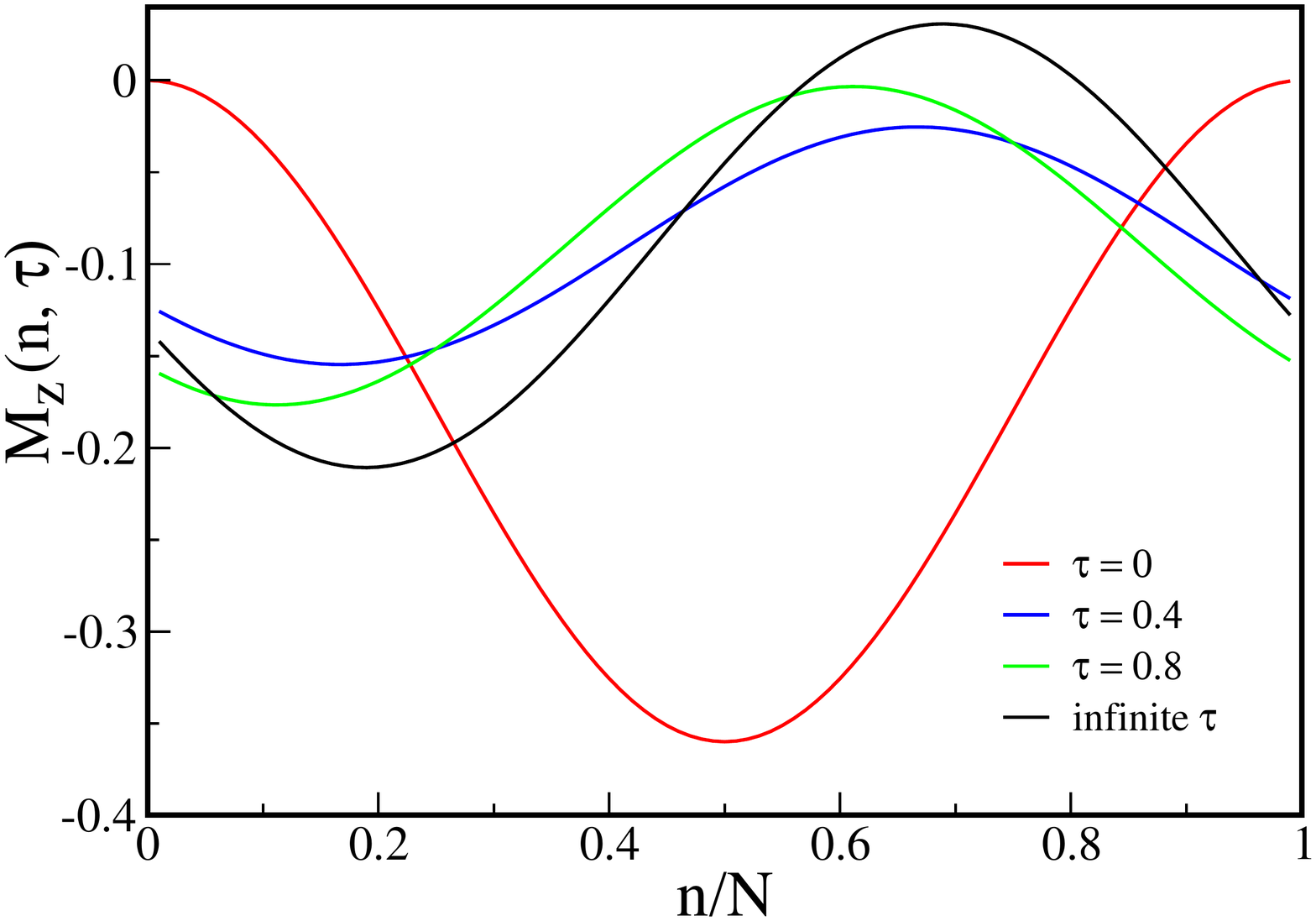}}
\caption{Transverse magnetization as a function of position $n$, starting from initial state (\ref{def_psi}) with 
$\alpha_k = 0.4$ for all $k$ and $\Gamma=0.5$. Here $\tau=t/N$.}
\label{mzvsn}
\end{subfigure}  
\end{figure}

Initial magnetization as derived from Eq. (\ref{M_int}) is given by
 \be M_z(n,t=0)=-\frac{1}{2}+2\alpha^2+(\beta^2-\alpha^2)\cos un \label{Mz-init}\ee 

and is shown in Fig. \ref{mz0-plot} for various $\alpha$. Moreover, Eqs. (\ref{M_int}) and (\ref{Mz-init}) indicate that the magnetization will be a function of position at all time (Fig. \ref{mzvsn}). \\

\begin{figure*}
{\includegraphics[clip,width=20cm, angle=0]{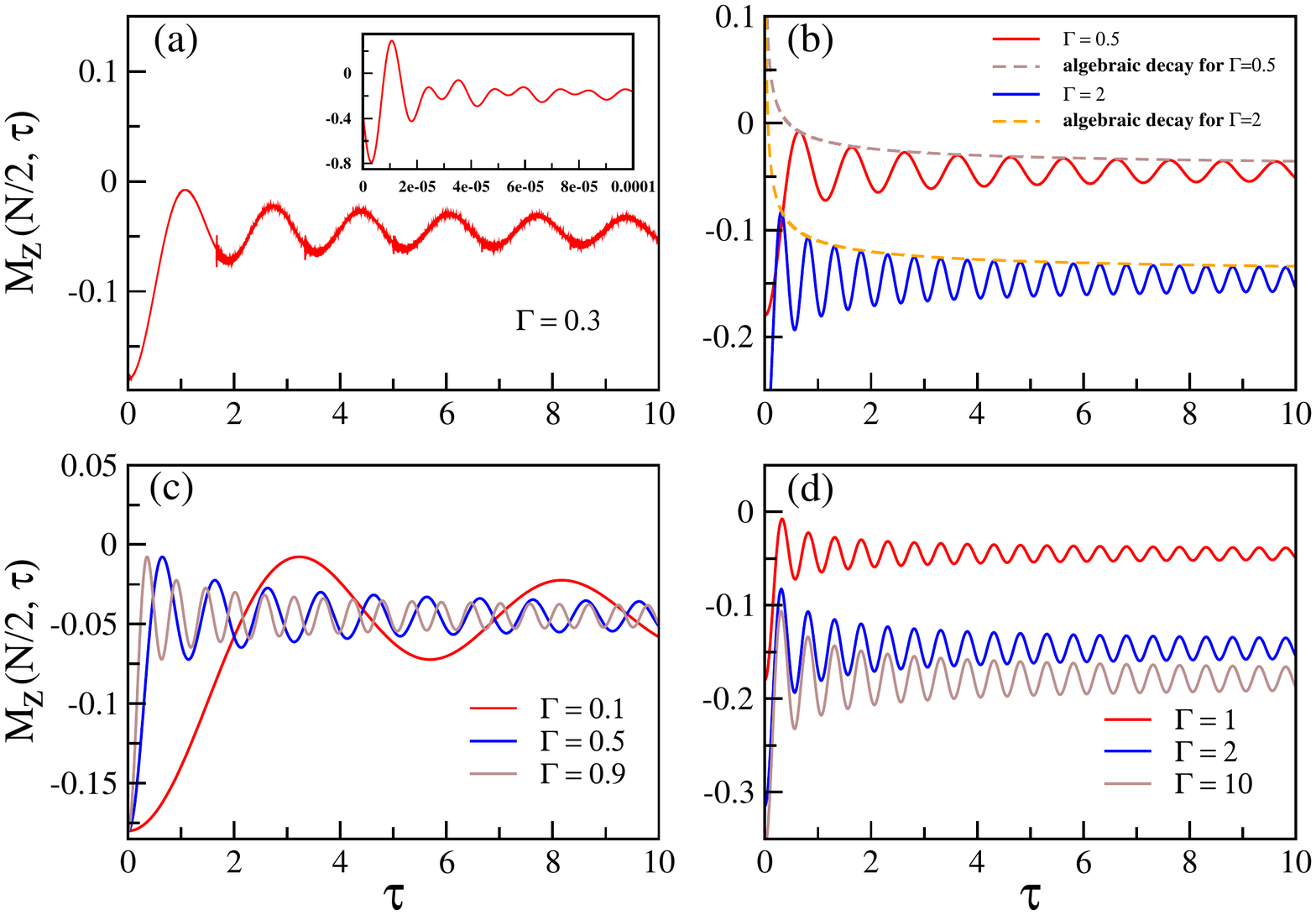}}
\caption{Plot of $M_z$ at $n=N/2$ against $\tau\;(=t/N)$, starting from initial state (\ref{def_psi}) with $\alpha_k =0.4$ for all $k$. $M_z$ is obtained by numerical integration of Eq. (\ref{M_int}) for $N=10^5$.
(a) $M_z$ vs $\tau$ alongwith small undulations due to time scale $t$. Variation of $M_z$ for very small change of $\tau$ given as inset shows oscillations in small timescale. (b) Oscillations in large timescale are shown in $M_z$ vs $\tau$ plot for $\Gamma=0.5$ and $\Gamma=2$. The envelopes of oscillations decay as $\tau^{-\frac{1}{2}}$ both for $\Gamma<1$ and $\Gamma>1$. (c) Oscillations in the region $\Gamma<1$, where the frequency increases monotonically with $\Gamma$. (d) The frequency of the same becomes constant for $\Gamma\geq 1$.}
\label{mzvst}
\end{figure*}

 The temporal behavior of transverse magnetization at a given site is shown in Fig. \ref{mzvst} for different values of external field. It exhibits oscillatory behavior with the envelope decaying algebraically in the large time limit. It is evident that Eq. (\ref{M_int}) has two time-dependent parts, $m_k(t)$ and $m_k(t/N)$ of two separate timescales. The former part gives tiny undulations (Fig. \ref{mzvst}(a)) and the other is responsible to produce larger oscillation with algebraic decay of envelope and dominates in the large time limit. Hence, for $t \gg 1$ and $t/N \gg 1$, we have
 \begin{eqnarray}
 M_z(n,\tau)&\approx & M_{\infty}(n)  \nonumber \\
 && + \frac{1}{\pi}\int_0^\pi [\; \frac{1}{2}(\cos^4\theta_k + \sin^4\theta_k) \nonumber \\
 && [2(\beta^2-\alpha^2)\cos un + \cos 2kn] \cos 2\pi\lambda'_k\frac{t}{N} \nonumber \\
 && -\; 2\alpha\beta\cos 2\theta_k \sin 2kn \sin 2\pi\lambda'_k\frac{t}{N} \;]\;dk \nonumber \\
&\approx & M_{\infty} + \frac{1}{\pi}\int_0^\pi \mathit{g}_k(n)\cos(\omega_k\tau + \phi_k(n))\;dk \nonumber \\
 \label{Mz-tau}
 \end{eqnarray}
where $M_{\infty} = -\frac{1}{2} + \frac{1}{\pi}\int_0^\pi \mathcal{M}_k \;dk$; $\omega_k = 2\pi\lambda'_k = 2\pi (d\lambda_k/dk)$ and $\tau=t/N$. The quantities $g_k(n)$, $\phi_k(n)$ can be written in terms of $\theta$ and $k$ from the last equation, but the explicit form is not necessary for our calculations below. We call the $\tau$-independent term $M_{\infty}(n) $ because it will give us the magnetization at $\tau\to\infty$ which we shall discuss later.

We now observe the behavior of $M_z$ for sufficiently large $\tau$ following the procedure in ref. \cite{SB1}. For large $\tau$, the quantity $(\omega_k \tau + \phi_k(n) )$ will be large so that its cosine will fluctuate very rapidly and vanish on integration unless $\omega_k$ is very small. Hence the region where $\omega_k$ is minimum with respect to $k$ will only contribute to the integral. This minima is found to occur at 
\be k = k_0 =\left\{ \begin{array}{rl}
      \cos^{-1}(- \Gamma)  & \mbox{for $|\Gamma| < 1$}\\
     \cos^{-1}( - \frac{1}{\Gamma}) & \mbox{for $|\Gamma| > 1$} \end{array} \right. \label{k0} \ee
      
Obviously $\dfrac{\pi}{2}\leq k_0 \leq\pi$. It is hence sufficient to integrate the oscillatory term over a small region ${k_0} - \epsilon< k < {k_0} + \epsilon$ ($\epsilon$ becoming smaller and smaller with increasing $\tau$) where $\mathit{g}_k(n)$ and $\phi_k(n)$ do not vary appreciably.  Expanding $\omega_k$ about $k=k_0$,
\be  \omega_k = \omega_{k_0} + \frac{1}{2}F(k-k_0)^2   \label{omega-Taylor}        \ee
(where $F = (d^2\omega_k / dk^2)_{k=k_0}$)
we get the expression for magnetization at large $\tau$ as,
\begin{eqnarray}
M_z(n,\tau) &\approx& M_{\infty} + \mathit{g}_{k_0}(n) \cos\left(\omega_{k_0}\tau + \phi_{k_0}(n)\right)\nonumber \\
&&\sqrt{\frac{2}{F\tau}}\int_{- \epsilon\sqrt{\frac{F\tau}{2}}}^{+ \epsilon\sqrt{\frac{C\tau}{2}}} \cos y^2 \;dy \nonumber \\ 
&\approx& M_{\infty} + \mathit{g}_{k_0}(n) \sqrt{\frac{\pi}{F}}  \frac{1}{\sqrt{\tau}}\cos\left(\omega_{k_0}\tau + \phi_{k_0}(n)\right) \nonumber \\
\label{M4} 
\end{eqnarray}
where $y = \sqrt{\frac{F\tau}{2}}(k - k_0 )$. Thus it is evident that for the chosen initial configuration, the transverse magnetization always exhibits $\tau^{-\frac{1}{2}}$ decay in the envelope of oscillation irrespective of the magnitude of the transverse field. Such decay was found in ref. \cite{SB1} also. \\

As shown in Fig. \ref{mzvst}(c) and (d), the frequency of oscillation varies monotonically with $\Gamma$ in the ferromagnetic phase and becomes independent of $\Gamma$ in the paramagnetic phase. It can be explained analytically. We note from Eq. (\ref{k0}) that $k_0$ changes from $\cos^{-1}(-\Gamma)$ to $\cos^{-1}(-\frac{1}{\Gamma})$ as $\Gamma$ crosses the critical point. This means that the frequency in the large $\tau$ limit behaves as 
\be |\omega_{k}| = 2\pi\left|\frac{d\lambda_k}{dk}\right| =\left\{ \begin{array}{rl}
      4\pi \Gamma & \mbox{for $|\Gamma| < 1$}\\
      4\pi & \mbox{for $|\Gamma| > 1$} \end{array} \right. \label{omegak0} \ee
The significance of this nonanalytic behavior of frequency is that it coincides with the occurrence of order-disorder quantum phase transition of the system.\\

The behavior at $\tau\to\infty$ (and hence $t\to\infty$ as well) can be obtained from Eq. (\ref{M_int}) by performing the integrations involved. For $\Gamma < 1$
\begin{eqnarray} && M_{\infty} (n)  = -\frac{1}{2}  + 2\alpha^2 -   \frac{\sin \, un}{4\pi} \left[  \frac{2}{\Gamma} + \frac{\Gamma^2  - 1}{\Gamma^2} \log \left|\frac{\Gamma+1}{\Gamma-1}\right| \right] \nonumber \\
&& + \frac{\beta^2 - \alpha^2}{4} \left( 2  -  \cos \, un \right)
 + \frac{\Gamma^{2n}}{16} \left(1 - \frac{1}{\Gamma^2}\right)  \label{Mz-inf1}  \end{eqnarray}
 and for $\Gamma < 1$,
 \begin{eqnarray} && M_{\infty} (n)  = -\frac{1}{2}  + 2\alpha^2 -   \frac{\sin \, un}{4\pi} \left[  \frac{2}{\Gamma} + \frac{\Gamma^2  - 1}{\Gamma^2} \log \left|\frac{\Gamma+1}{\Gamma-1}\right| \right] \nonumber \\
&& + \frac{\beta^2 - \alpha^2}{4\Gamma^2} \left( 2  -  \cos \, un \right)
 - \frac{1}{16\Gamma^{2n}} \left(1 - \frac{1}{\Gamma^2}\right).   \label{Mz-inf2}  \end{eqnarray}

The persistence of inhomogeneity of magnetization at $\tau\to\infty$ is evident. Another important aspect of the long-time behaviour is that the quantity $M_{\infty}(n)$ shows different behavior as a function of $\Gamma$ in ferromagnetic and paramagnetic region and is non-analytic at the critical point. Such behaviour originates from two integrals involved in the calculation of $M_{\infty} (n)$. They can be written as a contour integral over the unit circle and the integrand has poles at $z=\Gamma$ and $z=1/\Gamma$. For $\Gamma < 1$, the former pole is inside the unit circle and for $\Gamma > 1$, the latter is inside. Such swapping of poles that leads to a change in functional behavior and to non-analyticity has also been observed elsewhere \cite{SB2} in slightly different context.

\begin{figure}
{\includegraphics[clip,width=10cm, angle=0]{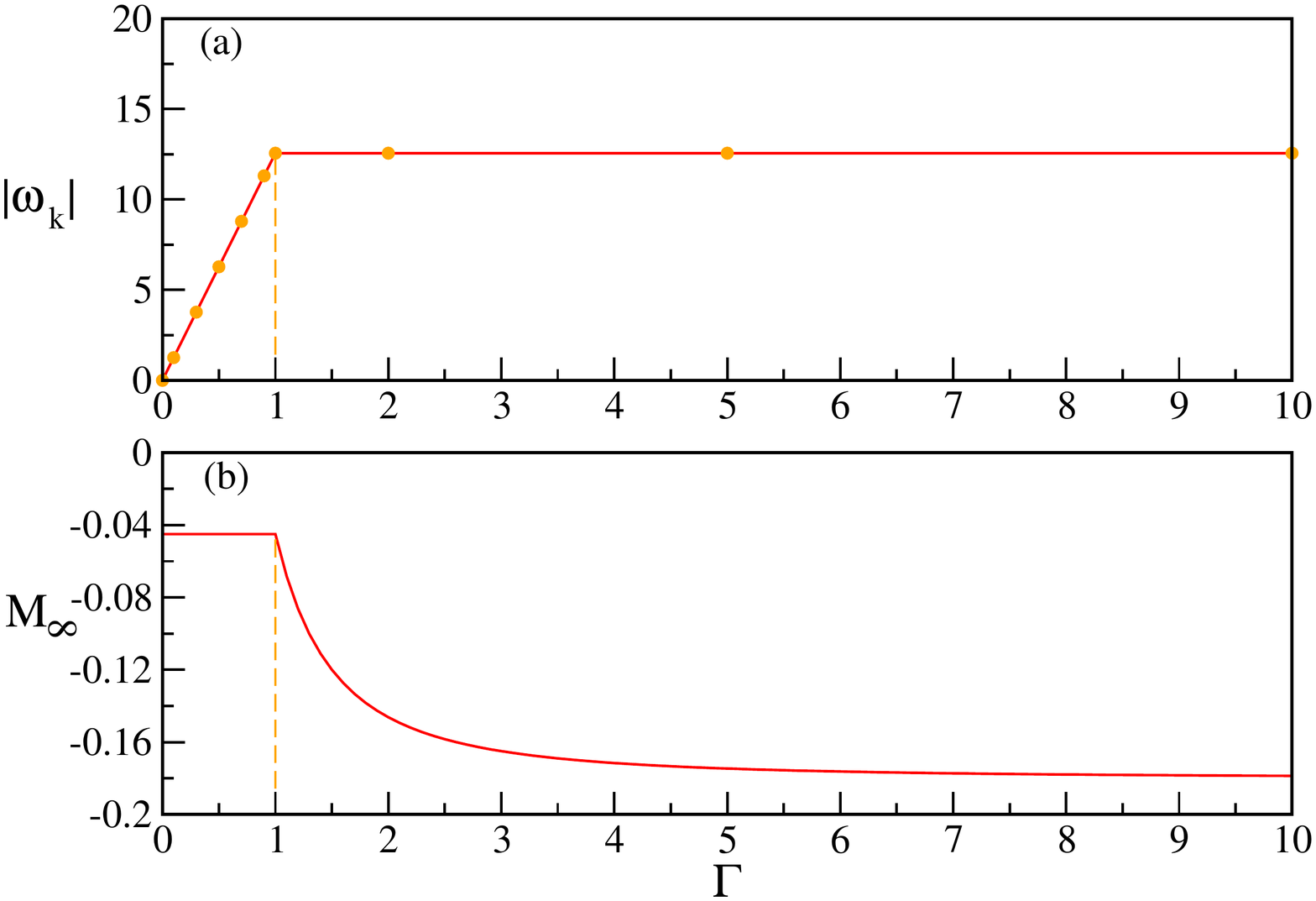}}
\caption{Signature of criticality : 
(a) $|\omega_{k}|$ vs $\Gamma$ plot \& (b) $M_{\infty}$ vs $\Gamma$ plot, for  $\frac{n}{N}=0.5$ with the initial state (\ref{def_psi}) taking $\alpha = 0.4$ for all $k$.}
\label{criticality}
\end{figure}

\section{Conclusion}

We have shown analytically that in a transverse Ising chain at zero temperature, a state can be constructed in the product structure in momentum space so that it produces spatial variation in transverse magnetization. Such inhomogeneity gives rise to transport of transverse magnetization from one site to another as a result of which both spatial and temporal variations take place even in presence of a homogeneous and constant (in time) external transverse field. At each site the magnetization evolves in an oscillatory manner with the envelope decaying algebraically with exponent $\frac{1}{2}$. The oscillation and decay however do not lead to homogeneous magnetization because the odd-occupation states have zero energy and their coefficients, which are the prime factor for inhomogeneity, remain nonzero forever. However, the retention of inhomogeneity has a deeper significance as it is connected with the integrability of the system. Being an integrable one, the local conservation laws possibly prohibit the system from getting a homogeneous magnetization profile. A non-integrable system, for the same reason, is expected to exhibit the opposite. Dissipative transport found in non-integrable system \cite{Meisner} also gives an indication to such behavior.

The exponent of decay is independent of the field i.e., it remains unaltered in ferromagnetic and paramagnetic phases. The characteristic change is shown by the frequency of oscillation at long time and the final magnetization at each site. They show different behavior in ferromagnetic and paramagnetic phases. 

Thus we find no signature of criticality in the exponent of decay whereas two new quantities are found to bear it. Moreover, starting with such configuration, the inhomogeneity in transverse magnetization cannot be removed by the external field. Both the phenomenon are counterintuitive to the prevailing ideas because the scenario may be thought of as a quench from an initial state and the preceding works on quench dynamics report the change of the exponent of decay in different phases \cite{Suzuki} and quantum many body system starting from any inhomogeneous magnetization is generally expected to attain homogeneity under influence of strong external parameter. What happens in our case is that the presence of odd-occupation states becomes the important controlling factor in dynamics. Although the initial configuration we have worked with contains real and $k$-independent values of the coefficients, it is important to note that, the principal observations, viz. $\tau^{-1/2}$ decay of envelope and behavioral change of $|\omega_k|$ and $M_{\infty}$ around quantum critical point are valid for arbitrary phases as well as modulus of $\alpha_k$  or $\beta_k$. Thus, in spite of the fact that the Hamiltonian is integrable, we observe a generic behaviour for a wide range of initial states, namely the ones satisfying Eq. (\ref{zero}).\\

A question arises regarding other observables. Integrable systems have been reported to have some non-local observables which show typically different behaviors including even thermalization in course of quench dynamics \cite{Suzuki}. Now in our system, with this initial configuration, how do non-local observables behave? Can thermalization or field-dependent decay exponent be found out in any of them? How do the non-integrable systems behave starting from such inhomogeneity? Does the prediction on attaining homogeneity become true for them? Search for answers of many such open questions may thus be quite interesting to investigate in future. \\

\begin{acknowledgement}
The authors are grateful to Abhishek Dhar and Arnab Das for valuable comments. SB acknowledges financial support under MRP scheme from UGC (India) with sanction no. F.PSW-043/15-16(ERO).
\end{acknowledgement}


\end{document}